\documentclass[conference]{IEEEtran}
\IEEEoverridecommandlockouts
\usepackage[numbers]{natbib}
\usepackage{amsmath,amssymb,amsfonts}
\usepackage{graphicx}
\usepackage{textcomp}
\usepackage{xcolor}
\usepackage{braket}
\usepackage{physics}
\usepackage{titlesec}
\usepackage{listings}
\usepackage[noend]{algpseudocode}
\usepackage{algorithm}
\usepackage{hyperref}
\usepackage{url}
\usepackage{xcolor}
\usepackage{color}
\usepackage{xspace}
\usepackage{multicol}
\usepackage{tabularx}
\usepackage{graphicx}
\usepackage{lscape}
\usepackage{tikz}
\usetikzlibrary{quantikz2}
\usepackage{multirow}
\usepackage{hyperref}

\hypersetup{
           breaklinks=true
        }
\usepackage{breakurl}

\definecolor{backcolor}{rgb}{0.95,0.95,0.92}
\definecolor{codegray}{rgb}{0.5,0.5,0.5}
\definecolor{codegreen}{rgb}{0,0.6,0}
\definecolor{codeblue}{rgb}{0.015,0.015,0.66}
\definecolor{codepurple}{rgb}{0.58,0,0.82}
\definecolor{codered}{rgb}{1.0, 0.0, 0}

\def\BibTeX{{\rm B\kern-.05em{\sc i\kern-.025em b}\kern-.08em
    T\kern-.1667em\lower.7ex\hbox{E}\kern-.125emX}}
\begin{document}

\title{Quantum Testing in the Wild: A Case Study with Qiskit Algorithms}

\author{
\IEEEauthorblockN{%
    Neilson C. L. Ramalho\IEEEauthorrefmark{1},
    Erico A. da Silva\IEEEauthorrefmark{1}, 
    Higor A. de Souza\IEEEauthorrefmark{2}, 
    Marcos Lordello Chaim\IEEEauthorrefmark{1}
    }
\IEEEauthorblockA{%
    \IEEEauthorrefmark{1}School of Arts, Sciences, and Humanities -- University of São Paulo, São Paulo, SP, Brazil\\
    \IEEEauthorrefmark{2}Department of Computing -- São Paulo State University, Bauru, SP, Brazil\\
    neilson@usp.br, augusto.ericosilva@usp.br, higor.amario@unesp.br, chaim@usp.br
}
}

\maketitle

\begin{abstract}
Although classical computing has excelled in a wide range of applications, there remain problems that push the limits of its capabilities, especially in fields like cryptography, optimization, and materials science. Quantum computing introduces a new computational paradigm, based on principles of superposition and entanglement to explore solutions beyond the capabilities of classical computation. With the increasing interest in the field, there are challenges and opportunities for academics and practitioners in terms of software engineering practices, particularly in testing quantum programs. This paper presents an empirical study of testing patterns in quantum algorithms. We analyzed all the tests handling quantum aspects of the implementations in the Qiskit Algorithms library and identified seven distinct patterns that make use of (1) fixed seeds for algorithms based on random elements; (2) deterministic oracles; (3) precise and approximate assertions; (4) Data-Driven Testing (DDT); (5) functional testing; (6) testing for intermediate parts of the algorithms being tested; and (7) equivalence checking for quantum circuits. Our results show a prevalence of classical testing techniques to test the quantum-related elements of the library, while recent advances from the research community have yet to achieve wide adoption among practitioners.
\end{abstract}

\begin{IEEEkeywords}
Quantum Software Engineering, Quantum Software Testing, Test Patterns, Empirical Study
\end{IEEEkeywords}

\section{Introduction}
Quantum computing has attracted the attention of both industry and academia in recent years, especially due to its capabilities to solve problems previously considered unmanageable for classical computers in areas like molecular simulations, cybersecurity, finance, and logistics. The main concept behind a quantum computer is to leverage the specific properties described by quantum mechanics to perform computation \cite{Hidary2019Quantum}.

The intersection between quantum mechanics and computation brings new challenges to classical software engineering practices, especially in handling concepts like superposition, entanglement, interference, and the probabilistic nature of quantum computers. While classical software engineering has built a robust theoretical and practical basis over recent decades, such practices for quantum computing are still emerging. Ensuring the correct functionality of quantum applications requires creating testing frameworks that support their development from theoretical and textbook algorithms to practical solutions addressing real-world problems. The research community has been actively adapting classical strategies and developing new testing techniques for quantum programs \cite{ramalho2024testingdebuggingquantumprograms}. In parallel, practitioners have started using quantum-specific frameworks and programming languages to implement quantum programs (QPs) to solve real-world problems. These QPs are inherently hybrid, with classical elements acting as integral parts of broader solutions rather than as isolated entities.

In this paper, we analyzed how practitioners apply testing techniques to the implementations of quantum algorithms. We manually examined the tests created for Qiskit Algorithms, a library of quantum algorithms built on Qiskit, designed to run on near-term quantum devices with shallow-depth circuits \cite{qiskit2024}. We went through the code of every single test method and classified it as either classical (basic validation and classical mathematical supporting functions, for instance) or quantum (tests for quantum algorithms or their properties). With this classification, we filtered out those that were purely classical tests and focused our analysis on the tests that address quantum-related concepts. The experimental dataset with the results and scripts for file pre-processing are available at GitHub \footnote{\url{https://github.com/saeg/saner2025/}}.

Our analysis uncovers testing patterns that employ classical software testing techniques such as black-box testing, gray-box testing with intermediate result checks, exact and approximate assertions, controlled randomness with fixed seeds, and data-driven testing. We also observe that the tests primarily run on simulators and do not account for transpilation in any of the cases.

These results suggest that practitioners have not yet adopted recent testing techniques such as those documented in recent studies \cite{ramalho2024testingdebuggingquantumprograms, GarciaQSTStateArt,QSE_Landscapes_Horizons, fortunato2024verification, paltenghi2024surveytestinganalysisquantum}. Possible reasons for the lack of usage of techniques developed by the research community might include the difficulty of applying the proposed techniques in practice or lack of awareness of the quantum software developers of testing techniques.

In what follows, Section \ref{sec:background} presents Qiskit Algorithms and the testing concepts addressed in this work. The experiment setup is detailed in Section \ref{sec:experiment}, followed by the results in Section \ref{sec:results} and our discussion in Section \ref{sec:discussion}. Section \ref{sec:threats} shows the threats to the validity. We briefly describe the related work in Section \ref{sec:relatedwork} and conclude the paper with potential research directions in Section \ref{sec:conclusions}.

\section{Background}
\label{sec:background}
Qiskit Algorithms is a Python library that provides quantum algorithms for use on simulators and near-term quantum devices with shallow circuits. It also includes key building blocks like quantum circuit gradients and state fidelities, which are commonly used in applications such as variational optimization, time evolution, and quantum machine learning \cite{qiskit2024}. 
It contains the categories of amplitude amplifiers, amplitude estimators, eigensolvers, gradients, minimum eigensolvers, optimizers, phase estimators, state fidelities, and time evolvers, according to the tasks they perform.

Qiskit Algorithms was the framework chosen for this study because (1) it includes tests for each algorithm, unlike most QP libraries; (2) it has an active GitHub community with 70 contributors; (3) it is built on Qiskit, one of the most widely used quantum development frameworks today; and (4) the algorithms are not limited to circuits only, but offer a basic framework that enables their application to real-world problems. 
As of October 2024, the Qiskit Algorithms testing suite contains 587 tests, which use different classical testing techniques such as precise and approximate assertions, functional testing, and DDT. 

Precise assertions are those that compare objects with the requirement of strict matching, i.e., with no room for tolerance, rounding, or approximation. Examples of such assertions in the Python framework unittest \cite{python_unittest_2024} are \texttt{assertEqual(a, b)} (checks if a and b are strictly equal) and \texttt{assertTrue(x)} (checks if x is true). Approximate assertions consider a tolerance when comparing the expected result with the one returned from the module being tested, as for instance, unittest's assertion \texttt{assertAlmostEqual}. NumPy\footnote{\url{https://numpy.org/}} assertions (such as \texttt{numpy.testing.assert\_allclose}) are also used in the tests, as they can do checks on arrays.

Another important concept discussed in this work is 
Data-Driven Testing (DDT), which consists of having a test method being executed multiple times with varying inputs and expected outputs \cite{Ammann_Offutt_2016}. This strategy promotes maintainability and helps reduce errors, as the multiple executions for the different parameters can be consolidated in a single test method, thus avoiding multiple tests with different names. 

These concepts are used throughout the Qiskit Algorithms code in combination with general testing techniques such as functional and gray-box testing. Functional testing, or black-box testing, is concerned with testing the program under the perspective of the requirements, looking at it as a black box without looking into its structure. The term gray-box testing is used for test generation approaches that use both source code internals as well as external descriptions of the software \cite{khan2012comparative}.

\section{Experimentation Setup}
\label{sec:experiment}

The experiment consisted of analyzing the code from the Qiskit Algorithms framework to investigate which software testing techniques were applied to the test of QPs. The steps are summarized as follows.

(1) Separation of classical- and quantum-related test methods.
Qiskit Algorithms contains not only a collection of quantum algorithms but also the necessary infrastructure to apply these algorithms in practical scenarios. Each algorithm is encapsulated in classes that contain constructors, getters, setters, and utility and optimization-related methods. This step consisted of removing the classical-related testing methods and analyzing the quantum-related ones to identify which testing techniques were used. The quantum testing techniques considered in our analysis are based on recent studies \cite{ramalho2024testingdebuggingquantumprograms, fortunato2024verification, paltenghi2024surveytestinganalysisquantum}. The analysis was performed manually and independently by the first and second authors of this work. The cases in which the classification for the tests differed were brought for discussion by all authors, and finally a consensus was reached. Hybrid tests (i.e., those covering both classical and quantum aspects of the code) were classified as  quantum tests.

(2) Assertions analysis. Based on the list produced by the previous step, we analyzed each test method to identify which assertions were being used. To achieve this, we executed a Python script to go through every file and count the occurrences of any method call that starts with ``assert'', covering assertions from both unittest and \texttt{numpy.testing} packages.

\section{Results}
\label{sec:results}

The code analysis performed in Qiskit Algorithms uncover seven recurring patterns across all the tests for quantum-specific parts of the algorithms. These patterns use (1) fixed seeds for the algorithms based on random elements; (2)  deterministic oracles; (3) precise and approximate assertions; DDT; (5) functional testing; (6) gray-box testing for intermediate steps; and (7) equivalence checking for quantum circuits. These patterns can be illustrated by a few examples taken from the tests created for the Variational Quantum Eigensolver (VQE) algorithm and detailed in the next subsections. The VQE is a hybrid quantum-classical algorithm that uses a variational technique to find the minimum eigenvalue of a given Hamiltonian operator $H$ \cite{Hidary2019Quantum}.

\subsubsection{Fixed Seeds} 
The tests for the VQE algorithm are located in the \texttt{TestVQE} class, within the \texttt{test/minimum\_eigensolvers/test\_vqe.py} file. Listing \ref{alg:setup_vqe} presents the \texttt{setUp} method for this class. In line 3, a random seed is set using \texttt{self.seed = 50} to ensure reproducibility in the parts of the algorithm that contain randomness, such as the parameters initialization and the classical optimizer.

\begin{lstlisting}[language=Python, label={alg:setup_vqe},caption=Setup method for the test class TestVQE]
def setUp(self):
    super().setUp()
    self.seed = 50
    algorithm_globals.random_seed = self.seed
    self.h2_op = SparsePauliOp(
        ["II", "IZ", "ZI", "ZZ", "XX"],
        coeffs=[-1.052373245772859,0.39793742484318045,
            -0.39793742484318045,-0.01128010425623538,
            0.18093119978423156])
    self.h2_energy = -1.85727503
\end{lstlisting}

Setting a fixed seed reduces flakiness in tests, ensuring consistent behavior across multiple executions \cite{FlakinessQPs}. 
The seed is not directly used in the tests but applied to the global constant \texttt{algorithm\_globals.random\_seed} from the package \texttt{qiskit\_algorithms.utils}, which is then used to build the global constant object \texttt{algorithm\_globals} (instance of \texttt{QiskitAlgorithmGlobals()}). 
This class contains a property called \texttt{random} that is derived from the \texttt{np.random.Generator} class and is used to create different probabilistic distributions (e.g., \texttt{algorithm\_globals.random.uniform} and \texttt{algorithm\_globals.random.normal}). 
This object is used across some of the classical optimizers in the \texttt{qiskit\_algorithms.optimizers} package and in quantum-related routines (e.g., the \texttt{validate\_initial\_point} method in the \texttt{utils} package). 

\subsubsection{Deterministic Oracles} From lines 5 to 9, the Hamiltonian operator \texttt{self.h2\_op} for the hydrogen molecule (H$_2$) is defined using Pauli strings with corresponding coefficients. This operator represents the electronic structure of H$_2$ mapped to a two-qubit system:

\begin{itemize}
    \item \textit{Pauli Strings:} Each string, such as $II$, $IZ$, $ZI$, $ZZ$, and $XX$, specifies the Pauli operators applied to the two qubits. For example, $II$ applies the Identity operator to both qubits, while $IZ$ applies the Identity to the second qubit and the Pauli-Z operator to the first qubit.
    \item \textit{Coefficients:} The coefficients represent the contribution of each term to the total energy, with values like $-1.052$ and $0.397$, as shown in the Listing \ref{alg:setup_vqe}.
\end{itemize}

The variable \texttt{self.h2\_energy = -1.85727503} contains the known ground-state energy of the H$_2$ molecule. The VQE algorithm uses \texttt{self.h2\_op} to compute the minimum eigenvalue, which should be equal (given a certain tolerance) to this value ($-1.85727503$). By comparing the computed energy with \texttt{self.h2\_energy}, one can validate that the VQE algorithm worked as expected (see line 8 in Listing~\ref{alg:test_vqe}).

\begin{lstlisting}[language=Python, label={alg:test_vqe},caption=Testing the VQE algorithm using gradient primitive.]
@data(CG(), L_BFGS_B(), P_BFGS(), SLSQP(), TNC())
def test_with_gradient(self, optimizer):
    estimator = Estimator()
    vqe = VQE(estimator, self.ry_wavefunction, optimizer,
        gradient=ParamShiftEstimatorGradient(estimator),
    )
    result = vqe.compute_minimum_eigenvalue(operator=self.h2_op)
    self.assertAlmostEqual(result.eigenvalue.real, self.h2_energy, places=5)
\end{lstlisting}

\subsubsection{Functional Testing} 
The test validates the output of the VQE algorithm (e.g., the minimum eigenvalue) without inspecting the internal parts of the quantum circuits or the optimization process. This tests the correctness of the algorithm as a whole, treating it as a ``black box.'' In line 7 of Listing~\ref{alg:test_vqe}, the main method of the VQE algorithm is called with the operator for the $H_2$ molecule (\texttt{h2\_op}) defined in the \texttt{setUp} method (Listing~\ref{alg:setup_vqe}). The outcome of the method call is saved into the variable \texttt{result} and compared to the expected, pre-calculated value (\texttt{h2\_energy} variable) in the \texttt{assertAlmostEqual} method, with a tolerance of $10^5$ (places = 5).

\subsubsection{Gray-box Testing}

Listing \ref{alg:test_gradient_calculation} presents the \texttt{test\_gradient\_calculation} method for the \texttt{TestAdaptVQE} class. The test directly interacts with the internal gradient calculation process, checking the correct computation of commutators between operators. This low-level test is a form of gray-box testing as it verifies a specific internal method (\texttt{\_compute\_gradients}), providing visibility into the quantum-related operations before the full AdaptVQE algorithm is executed. The method itself is tested as a black box by comparing the computed gradient result with a manually calculated reference value. The test then checks the correctness of this intermediate result, which is needed for further iterations of the algorithm.

\begin{lstlisting}[language=Python, label={alg:test_gradient_calculation},caption=Test gradient calculation method in AdaptVQE.]
    def test_gradient_calculation(self):
        solver = VQE(Estimator(), QuantumCircuit(1), self.optimizer)
        calc = AdaptVQE(solver)
        calc._excitation_pool = [SparsePauliOp("X")]
        res = calc._compute_gradients(operator=SparsePauliOp("Y"), theta=[])
        # compare with manually computed reference value
        self.assertAlmostEqual(res[0][0], 2.0)
\end{lstlisting}

Another example of gray-box testing in Qiskit Algorithms is the test of the callback functions of the iterative, hybrid algorithms. For instance, the method \texttt{test\_callback} in the \texttt{TestVQD} class (file \texttt{test/eigensolvers/test\_vqd.py}) tests a callback mechanism that allows for the observation of intermediate states during optimization, which gives visibility into the system's internal state during execution and could support further metamorphic testing approaches if conditions or outputs were cross-checked.

For many of the analyzed test methods categorized as gray-box, Qiskit Algorithm's developers created complex input datasets to thoroughly exercise the important paths and conditions of the algorithms being tested. This need to understand the algorithm's internal functioning to generate appropriate test data was the decision criterion for classifying these tests as gray-box rather than black-box.

Of the 309 analyzed tests, 63 are classified as purely black-box (i.e., the test consists of calling a quantum-related routine and asserting the results against a pre-calculated value), and 80 are classified as gray-box (in which the test case and input data are prepared considering prior knowledge of the algorithm's internal structure). The remaining 166 tests are classical and do not involve quantum-related concepts.

\subsubsection{Classical Assertions}
Assertions such as \texttt{assertAlmostEqual} from the unittest Python framework or \texttt{np.testing.assert\_allclose} from NumPy are used extensively throughout the tests for the Qiskit Algorithm framework. As exemplified in line 8 of Listing \ref{alg:test_vqe}, the result from the \texttt{compute\_minimum\_eigenvalue} method might not strictly match the pre-calculated value expected as the response. However, with the \texttt{places} parameter, the developer can control the tolerance for the assertion and guarantee that deviations due to the randomness of the quantum algorithm or possible floating point issues can be properly handled. The results with the top five most frequent assertions are summarized in Table \ref{tab:assertions_summary}. 

\begin{table}[htbp]
    \centering
    \renewcommand{\arraystretch}{1.2} 
    \caption{Summary of Assertion Types and Occurrences}
    \label{tab:assertions_summary}
    \begin{tabular}{|l|r|r|}
        \hline
        \textbf{Assertion Name} & \textbf{Occurrences} & \textbf{Percentage} \\
        \hline
        assertEqual & 135 & 27.95\% \\
        \hline
        assertAlmostEqual & 128 & 26.50\% \\
        \hline
        assert\_allclose & 59 & 12.22\% \\
        \hline
        assertIsInstance & 38 & 7.87\% \\
        \hline
        assertTrue & 34 & 7.04\% \\
        \hline
    \end{tabular}
\end{table}

\subsubsection{Data-Driven Approach} 

As the VQE uses a classical optimizer, the test in Listing \ref{alg:setup_vqe} runs once for each optimizer passed in the \texttt{@data} parameter (\texttt{CG(), L\_BFGS\_B(), P\_BFGS(), SLSQP(), TNC()}). This shows that the test explores the hybrid behavior of the VQE algorithm, as the optimizers are classical routines. The use of DDT in this case is convenient, as the developer needs to implement a single test method that runs once for each passed parameter. 
This makes testing code more maintainable, for instance, when a new classical optimization routine is added to the project. Instead of creating a separate method, the developer needs to add the name of the new optimization routine in the \texttt{@data} parameter and the test will run for it as well. The DDT-related annotations (\texttt{@data} and \texttt{@idata}) are used in 133 test methods, accounting for 43\% of the total 309 tests in the suite.

\subsubsection{Equivalence checking for quantum circuits}
There are tests in which the developer compares the unitary matrix from two circuits to determine whether they are equivalent. For instance, the \texttt{TestBernoulli.test\_qae\_circuit} method (\texttt{test/test\_amplitude\_estimators.py} file) is designed to test the correctness of the circuit generated for the Amplitude Estimation algorithm, which uses a quantum circuit that is built and optimized to low qubit usage and reduced gate depth. This test manually constructs the amplitude estimation circuit and compares the resulting unitary matrix with the unitary matrix generated by the optimized version of the circuit that is implemented in the \texttt{AmplitudeEstimation} class. The goal of the test is to verify that the optimization preserves the correctness of the circuit.
This test verifies the circuit creation process for circuits with 2 to 5 qubits. Since it compares unitary matrices to determine equivalence, there will be scalability issues in case the number of qubits increases. 

\section{Discussion}
\label{sec:discussion}

Testing quantum code is a complex task. The developer needs to have a good understanding of the algorithm being tested and be able to identify preconditions, post-conditions, intermediate results, and edge cases. 

To handle the inherent complexity, our analysis of the test methods in Qiskit Algorithms shows that developers often opt for simpler, classical testing techniques when testing quantum algorithms, as summarized in Table \ref{tab:testing_techniques_summary}. The hybrid nature of these algorithms is evident from the distribution of testing techniques within the test suite. As expected, classical components still play a significant role, accounting for 54.69\% of the total tests. This predominance is explained by the fact that quantum algorithms require classical infrastructure to interface with the classical world (through input and output encoding, communication with classical optimizers, and structural code such as getters, setters, constructors, helper functions, and mathematical routines). For quantum-related parts, developers typically use either black-box testing (20.39\%) or gray-box testing (24.92\%), where knowledge of certain internal conditions and elements of the algorithms’ structure is used to design test data that exercise these specific components. 

\begin{table}[htbp]
    \centering
    \renewcommand{\arraystretch}{1.2} 
    \caption{Summary of Testing Techniques and Occurrences}
    \label{tab:testing_techniques_summary}
    \begin{tabular}{|l|r|r|}
        \hline
        \textbf{Testing Technique} & \textbf{Number of Tests} & \textbf{Percentage} \\
        \hline
        Black-box Testing & 63 & 20.39\% \\
        \hline
        Classical & 169 & 54.69\% \\
        \hline
        Gray-box Testing & 77 & 24.92\% \\
        \hline
    \end{tabular}
\end{table}

The use of functional testing combined with fixed seeds in Pseudo-Random Number Generators (PRNGs) and precise and approximate assertions is a predominant pattern in the test cases for the quantum-related parts of the library. The use of fixed seeds, for instance, has been listed by previous works \cite{FlakinessQPs} as a solution for flakiness in tests of QPs.

As for the usage of assertions, the data indicates that among all quantum-related test methods, \texttt{assertEquals} is the most frequently used assertion, accounting for nearly 30\% of all assertions. This result shows that although the probabilistic nature of QP is a concern for the research community, in practice it is handled by using known answers for certain parameters when executing the quantum algorithms and fixed seeds. For the cases in which floating point precision might be an issue, the solution is to use an approximate assertion with a certain precision, which explains  \texttt{assertAlmostEqual} as the second most used assertion, also accounting for almost 30\% of the total number of assertions in quantum-related tests.

DDT is a classical testing technique that is extensively used in the Qiskit Algorithms test methods. Our analysis shows that the use of DDT reduces test complexity, as it helps separate test logic from data and makes it easier to manage several combinations of input parameters and expected outputs.

In terms of gray-box testing, the approaches adopted in the tests are more related to the test of intermediate routines in the algorithms. There are no tests for branches, conditions, or individual statements. These intermediate routines are then tested using functional testing with a pre-calculated expected value using precise and approximate assertions. Some quantum algorithms allow for callback functions to be passed as arguments in the iterative part of the algorithm (VQE, for instance). For these algorithms, the tests check whether intermediate results (e.g., evaluation count, parameters, mean values) can be stored during the optimization process using a callback function. This testing approach explores the interface between the quantum and the classical parts of the algorithm.

The tests in Qiskit Algorithms with quantum-related elements are executed using either the statevector simulator or the Qasm simulator (shot-based). There are no interactions with real quantum computers, as access to quantum hardware is still limited. However, to better understand how these techniques perform in real-world scenarios, it would be valuable to run the tests on real quantum hardware. Approaches to make testing more realistic could include using recordings, similar to those employed in Azure Quantum tests \cite{azure_quantum_tests_readme}. In the Azure Quantum Python project, the testing infrastructure uses Python VCR \cite{vcrpy} to record HTTP calls against a live service. These recordings (or cassettes) are used to playback the responses that work as a mock of the live service. Transpilation is another topic not covered in the analyzed test suite, as no specific transpilation setup is defined in tests that depend on circuit simulation. This can be problematic as transpilation can not only adapt the circuit to the target architecture but it also optimizes such a circuit. While optimization often reduces gate count, it can also increase circuit depth and width, potentially adding more gates and even introducing issues like the Long Circuit smell in previously clean circuits \cite{Stefano2024}.

From the tests classified as having quantum features, we found no evidence of the testing techniques for QPs presented in recent works \cite{ramalho2024testingdebuggingquantumprograms, GarciaQSTStateArt,QSE_Landscapes_Horizons, fortunato2024verification, paltenghi2024surveytestinganalysisquantum}. This gap needs to be explored further, as it may indicate that practitioners are either unaware of recent advancements in QP testing or that these techniques are difficult to apply in practice. Another possibility is that academic research may face practical limitations in implementation, either due to a lack of ongoing maintenance or an inability to scale effectively for real-world applications.

Mutation testing of quantum programs is well-studied in research, offering frameworks for mutating quantum circuits \cite{ArcainiMuskit, MutationTestingRuiabreu2, gil2024qcrmutquantumcircuitrandom}. However, these techniques are currently not used to evaluate test suite quality for Qiskit Algorithms. This may occur because mutation operators focus on the circuit itself, while Qiskit Algorithm artifacts are designed to be components of larger applications where circuits are just part of the solution. For algorithms based on parameterized circuits, the circuit serves as a scaffold for parameter optimization \cite{Hidary2019Quantum}. Mutating these circuits during training may not reflect bugs that a developer would actually introduce.

\section{Threats to validity}
\label{sec:threats}
Our study has two main threats to its validity, namely, internal and external validity, since it is largely observational and does not rely on statistical analysis.

\textit{Internal validity}. The manual analysis of the testing techniques might be a process prone to errors as it depends on human judgment, which can introduce subjectivity and inconsistencies. However, the classification was initially performed by the first two authors and subsequently discussed by all authors. This process reduces the chances of mistakes. The process of looking for testing techniques consisted of analyzing each test case and classifying it as either quantum or classical.

\textit{External validity}. Qiskit Algorithms may not fully represent how the broader community applies quantum testing techniques. However, it is built on Qiskit, one of the most active and widely used frameworks for quantum programming development. The repository is licensed under Apache 2.0, remains active, and implements several fundamental building blocks of quantum applications, including Quantum Phase Estimation (QPE), Quantum Phase Amplification (QPA), Grover’s Algorithm, and Variational Quantum Eigensolvers (VQE).

\section{Related Work}
\label{sec:relatedwork}
Previous studies have analyzed quantum programs (QPs) to identify bug patterns \cite{ZhaoIdentifyingBugPatterns}, examining real-world bugs and bug fixes \cite{ComprehensiveStudyBugFixes, PaltenghiBugsInQP}, and focusing specifically on bugs within quantum machine learning applications \cite{zhao2023empirical}. Other approaches employ static and dynamic analysis to detect bugs in QPs \cite{QChecker, paltenghi2024analyzingLintQ} and investigate code smells \cite{TheSmellyEight}. To the best of our knowledge, our work is the only study focused on analyzing patterns in testing techniques applied to real QPs.

\section{Conclusions and Future Work}
\label{sec:conclusions}
Quantum computing has become a promising field due to its potential to tackle complex problems. As interest grows in quantum programming languages and tools, it is important to develop and refine techniques for testing quantum programs. 

This paper brought an overview of the main testing patterns used by practitioners to test the Qiskit Algorithms framework. We showed that, although the research community has started developing techniques to test different parts of a QP, in practice, developers continue using classical strategies to test quantum algorithms. 

Our results highlight the importance of filling the gap between academia and practitioners in terms of the testing strategies for QPs. On one hand, there are multiple active research groups worldwide developing techniques to test QPs, which are not used in practice, to the best of our knowledge. On the other hand, developers might be missing opportunities to apply the state-of-the-art in terms of testing methods for QPs and improve the overall quality of the artifacts they produce. 

In future work, we plan to extend this analysis to consider other programming languages such as Q\#, frameworks from Qiskit-Community GitHub repository\footnote{\url{https://github.com/qiskit-community}}, and frameworks such as PennyLane\footnote{\url{https://pennylane.ai/}} and Cirq\footnote{\url{https://quantumai.google/cirq}}. The idea is to check whether the patterns observed in Qiskit Algorithms are also present in these other frameworks and languages as well as identify any new patterns that may emerge. Another possible extension of this work is identifying opportunities in the frameworks' code to apply testing techniques developed by the research community. Future investigations could explore how easily these techniques can be applied in practice and the benefits they offer compared to current classical approaches.

\bibliographystyle{IEEEtranN}
\bibliography{references}
\end{document}